\documentclass{iau}
\usepackage{graphicx}

\title[Using tailed radio galaxies as environmental probes] 
{Using the morphology and magnetic fields of tailed radio galaxies as environmental probes}

\author[M. Johnston-Hollitt et al.]   
{
M. Johnston-Hollitt$^1$,
S. Dehghan$^1$ 
\and
L. Pratley$^1$
}

\affiliation{ 
$^1$School of Chemical \& Physical Sciences, Victoria University of Wellington, \\
PO Box 600, Wellington 6140, New Zealand \\ 
email: {\tt Melanie.Johnston-Hollitt@vuw.ac.nz} 
}

\pubyear{2014}
\volume{313}
\pagerange{xx}
\jname{Extragalactic jets from every angle}
\editors{F. Massaro, C.C. Cheung, E. Lopez, A. Siemiginowska, eds.}

\begin{document}

\maketitle

\begin{abstract}
Bent-tailed (BT) radio sources have long been known to trace over densities in the Universe up to $z \sim 1$ and there is increasing evidence this association persists out to redshifts of 2. The morphology of the jets in BT galaxies is primarily a function of the environment that they have resided in and so BTs provide invaluable clues as to their local conditions. Thus, not only can samples of BT galaxies be used as signposts of large-scale structure, but are also valuable for obtaining a statistical measurement of properties of the intra-cluster medium including the presence of cluster accretion shocks $\&$ winds, and as historical anemometers, preserving the dynamical history of their surroundings in their jets. We discuss the use of BTs to unveil large-scale structure and provide an example in which a BT was used to unlock the dynamical history of its host cluster. In addition to their use as density and dynamical indicators, BTs are useful probes of the magnetic field on their environment on scales which are inaccessible to other methods. Here we discuss a novel way in which a particular sub-class of BTs, the so-called `corkscrew' galaxies might further elucidate the coherence lengths of the magnetic fields in their vicinity. Given that BTs are estimated to make up a large population in next generation surveys we posit that the use of jets in this way could provide a unique source of environmental information for clusters and groups up to $z$ = 2.

\end{abstract}

\firstsection 
\section{Introduction}

Bent-Tailed (BT) radio sources are an intermediate population of radio galaxies between Fanaroff-Riley (FR) class I \& II sources. BTs with their bright, distorted jets and non-linear radio structure are readily distinguished from standard double radio galaxies, i.e.\ FRII sources. The complex radio structures of BTs are believed to be the result of a number of environmental effects. Firstly, the motion of the host galaxy through a dense medium generates a strong ram pressure on the jets, thus warping the structure. Whilst ram pressure is often responsible for the overall morphology of a BT source, other physical processes can also play an important role in the morphology of BTs on different scales. Firstly, density perturbations in the medium may induce buoyancy forces on jets which can further distort the structure. Precession of the host galaxies can further complicate the shape of the jets and finally, gravitational effects of nearby objects in interacting systems have been suggested to cause symmetric or asymmetric distortions in the radio morphology of BTs producing characteristic twists, bends and kinks in their tails. In some situations all these mechanism might be required to explain the shape of a particular tailed source. This implies that tailed sources are unique probes of a number of mechanism, over a range of physical scales from cluster-wide winds to binary galaxy pair gravitational interactions. By exploiting the fact that high galaxy concentrations are a general requirement for all the mechanisms mentioned above, and in fact, BTs are primarily found in galaxy clusters and groups, tailed radio sources have been suggested as a tool to identify overdensities both on the scale of clusters and to pinpoint the highest density regions within those clusters (Blanton et al., 2001 \& 2003; Mao et al. 2009, Mao et al., 2010).

As the morphology of BTs is principally considered as a function of the local environment, it is possible to measure the physical properties of the host structure, such as the density of the medium (Freeland et al., 2008), the dynamical status (Pfrommer \& Jones, 2011), and the magnetic field strength of the surrounding medium (e.g. Pratley et al. 2013 and references therein). Here we consider the use of tailed radio galaxies as both markers of overdense regions and probes of the physical conditions therein. In particular, we consider BTs as probes to cluster weather and magnetic fields.

\begin{figure}
\begin{center}
\includegraphics[width=13.8cm]{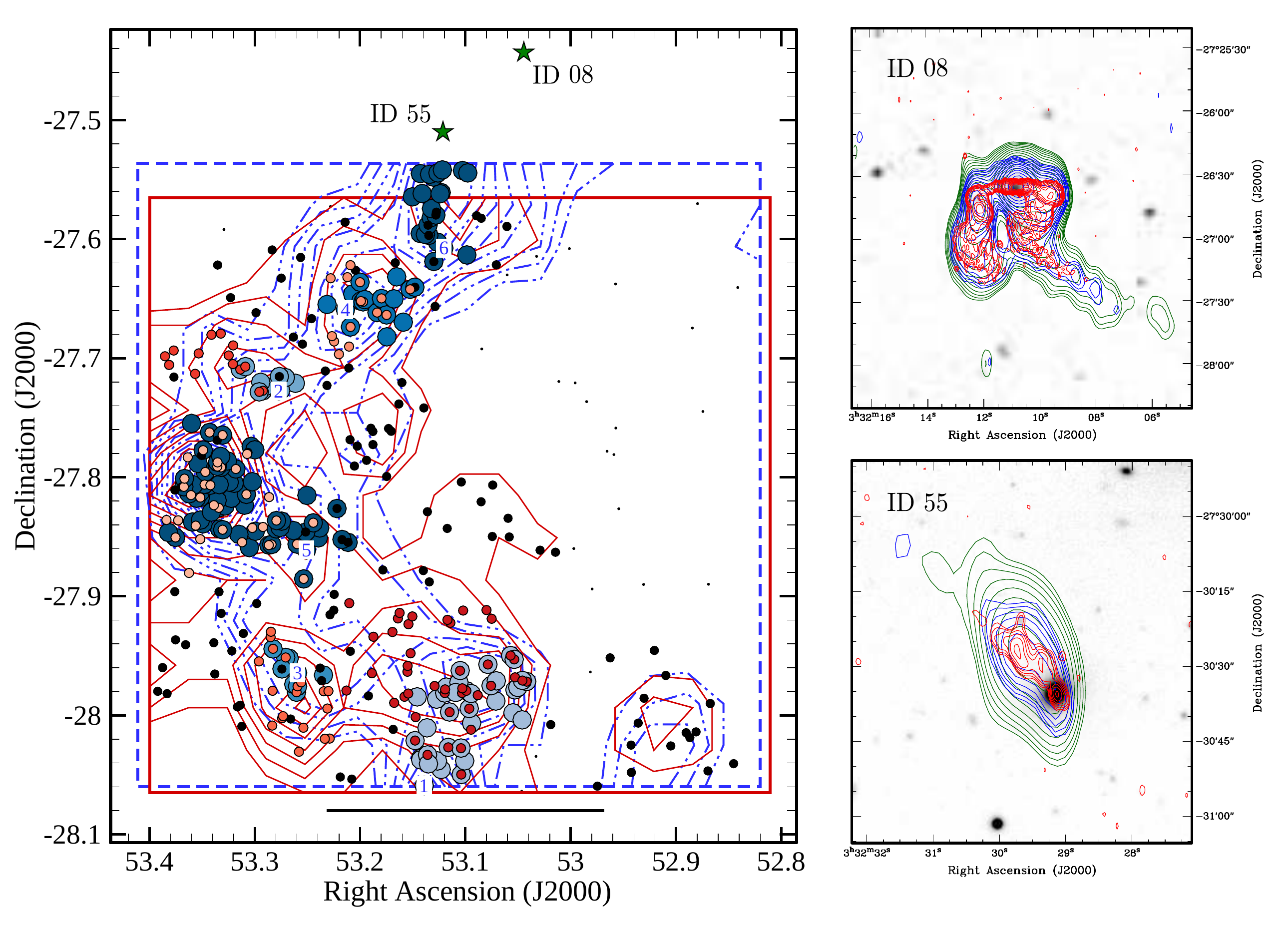}
 \caption{Detection of large scale structure in the CDFS and its relation to the location of BTs. Left: Large scale structure detected in a particular redshift slice ($0.100 \leq z \leq 0.175$) of the CDFS taken from Dehghan \& Johnston-Hollitt (2014). Spectroscopic and photometric groups detected in this region are respectively shown by the small dots with the red colour scheme and the large dots with the blue colour scheme. Similarly the red solid contours and blue dot-dashed contours given the spectroscopically and photometrically derived galaxy surface densities, respectively.  The black line at the bottom of the plot represents 2 Mpc in angular extent at the average redshift of this region $z \sim 0.13$. Green stars give the location of two tailed galaxies in this region believed to be associated with the large-scale structure.  Right: Radio observations of the BTs marked by green stars on the left panel that appear in the same redshift slice. Green and blue contours show the 1.4 GHz ATLAS first and third data release at 7$^{\prime\prime}$ while red contours present the 1.4 GHz VLA data at 2$^{\prime\prime}$ resolution. Background images are  extracted from GaBoDS. Radio contours start at $3\sigma$ and increase at intervals of $\sqrt{2}$. The IDs (08 \& 55) correspond to the catalogue numbers taken from the catalogue of all BT sources in the extended Chandra Deep Field-South (Dehghan et al. 2014).}
  \label{fig:LSS-HT}
\end{center}
\end{figure}

\section{Bent-Tailed radio galaxies as probes of large scale structure}

\begin{figure}
\begin{center}
\includegraphics[width=13.7cm]{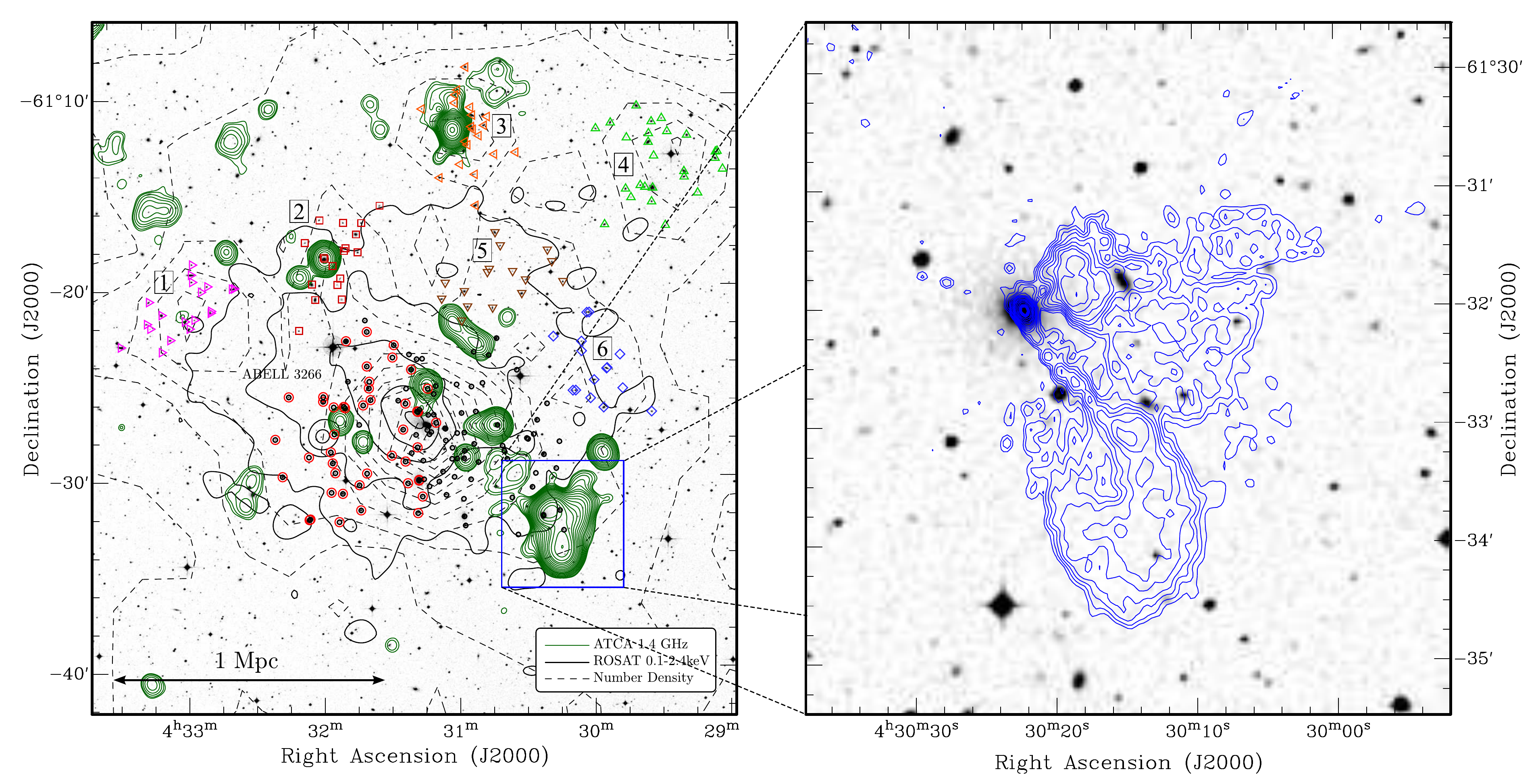}
 \caption{Multiwavelength information on the galaxy cluster A3266 (taken from Dehghan 2014). Left: Coloured symbols give the spatial distributions of the spectroscopically detected sub-structures (filaments and groups) in the cluster, overlaid on DSS optical image. The members of the main component of the cluster core are shown with filled black circles, while substructure in the core is shown via red circles. Black dashed contours provide the optical density for cluster members and black solid contours give the soft-band X-ray (0.1-2.4 keV) emission from XMM-Newton. The galaxy density contours start at the local $3\sigma$ rms level and then increase by steps of 11 galaxies per Mpc$^2$. The scale line represents 1 Mpc angular extent at the mean redshift of the cluster core. Green contours show the 1.4 GHz ATCA radio emission convolved to 15$^{\prime\prime}$ resolution and the head-tailed galaxy, PKS J0430-6132, is denoted with a blue frame.  Right: ATCA 1.4 GHz contours of the tailed radio galaxy, PKS J0430-6132, at full resolution (6$^{\prime\prime}$) overlaid on the DSS optical image. Contours start at $3\sigma$, 0.76 mJy beam$^{-1}$ and increase by a factor of $\sqrt{2}$. In order to take this shape we believe the tailed galaxy was moving to the southwest and then made a rapid turn back to the cluster core during the second core passage of the subgroup marked by red circles in the left panel.}
  \label{fig:A3266}
\end{center}
\end{figure}

The association between BTs and galaxy clusters and groups, which persists up to at least redshift 1 (Blanton et al., 2003), suggests a promising method for tracing large scale structures, both in the local and distant universe. Furthermore, there is growing evidence that such an association is likely to hold up to redshift 2 (Dehghan et al., 2011 \& 2014), when the first clusters formed. If this is shown to be the case BTs in wide-area radio surveys of the future could become a powerful tool for detection of clusters. Two approaches have been taken to date to test the reliability of this method for detection of overdensities, the first is the large-scale study of BT sources in the FIRST survey (Blanton et al. 2001, Wing \& Blanton 2011 and Blanton et al. this volume) in which BTs are determined from the radio survey and then followed up to search for overdense regions. The second approach focuses on 
cross comparisons of large scale structures and radio images in legacy fields such as the Chandra Deep Field-South (CDFS), where substantial multi-frequency and spectroscopic data are available. Such a comparison between the detected structures at $z \sim 0.13$ (Dehghan \& Johnston-Hollitt, 2014) and BTs found at the same redshift (Dehghan et al., 2014) in the CDFS is shown in Figure \ref{fig:LSS-HT}. BTs, labelled ID 08 \& 55 (shown by green stars in the left panel of Figure \ref{fig:LSS-HT}), reside in what appears to be a continuation of the spectroscopically detected arc-shaped large scale structure. This is suggestive, but not definitive evidence, that BTs not only reside in cluster, but also in large scale structures in general. These two approaches are complementary and similar studies on both wide-area radio surveys and other legacy fields will shed more light on the correlation between BTs and overdensities in the Universe, particularly as a function of redshift.

\section{Using Bent-Tailed radio galaxies to characterize environmental conditions}

When BTs are detected in clusters they can be readily used to probe a range of environmental conditions including the presence of accretion shocks (Pfrommer \& Jones 2011), cluster densities (Freedland et al. 2008) or velocity flows (Douglass et al. 2011) in the Intra-Cluster Medium (ICM). Additionally, detailed modelling of tailed radio galaxies has been used to not only reconstruct the morphology of the tails but also provide information on the dynamical history of the host cluster by acting as anemometers for past cluster winds (e.g., Pratley et al. 2013, 2015). Recently, we have used the prominent BT, PKS J0430-6132, in A3266 to break the degeneracy between possible merger stages. Detailed spectroscopic optical analysis confirmed the merger axis (northeast to southwest) and presence of substructure in the cluster core (Dehghan 2014, Dehghan et al. in prep) but was unable to distinguish between on-going in-fall of a sub-cluster from the southwest with a single core passage or alternatively a merger from the northeast on the second core crossing with associated sloshing. The morphology of the BT seen in our detailed radio imaging of the southwest of A3266 (Figure \ref{fig:A3266}; right panel), combined with modelling pointed to the latter scenario in which the tailed galaxy was travelling to the southwest then made a rapid reversal back towards the cluster core as the second core crossing of the original in-falling group occurred. Full details will be presented in Dehghan et al. (in prep). This highlights how even simple BTs can also be used to constrain the dynamical history of a complex cluster system.

\begin{figure}
\begin{center}
\includegraphics[width=13.5cm]{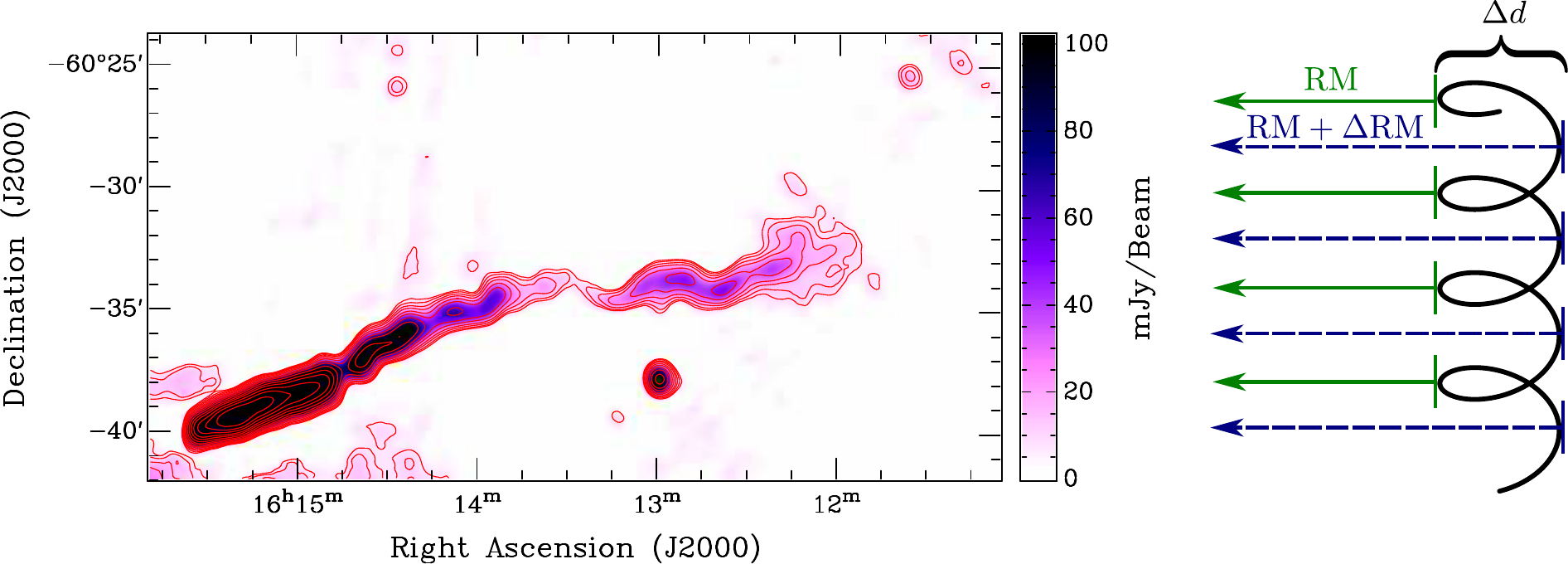}
 \caption{Left: The `corkscrew' galaxy, ESO 137-G 007, in the Norma cluster at 843 MHz from the Sydney University Molonglo Sky Survey (SUMSS). Note the tightly wound helical path of the radio jet. Right: Depiction of the different path lengths, $\Delta d$, through the ICM for emission from the near and far side of the helical jet. Polarised emission from the nearside of the jet (green solid arrows) will probe a fraction of the ICM which is less than that of emission from the far side of the jet (blue dashed arrows), as a result an additional rotation measure ($\Delta$RM) component will be present. Examination of the difference between RMs for the near and far sides of the helical jets probes the magnetic field on scales of the width of the helical path, $\Delta d$.}
  \label{fig:helix}
\end{center}
\end{figure}

\section{Bent-Tailed radio galaxies as probes of the intra-cluster magnetic field}

The use of BTs to probe the magnetic field in galaxy clusters has had a long history, particularly as polarized sources for Faraday rotation (Pratley et al., 2013 and references therein). Traditionally the large, resolved tails have been used as Faraday screens to probe the coherence length of the cluster magnetic field on smaller scales than are accessible through other means. Using this method provides information on the average coherence length of the magnetic field in the plane of the sky, over the ICM between the tails and the edge of the cluster. Since rotation measures are susceptible to all magnetic media along the line-of-sight, using the resolved tails of galaxies gives an ensemble average of the coherence scale in the cluster field along with that of any intervening field such as that of the Milky Way. Although such techniques have suggested cluster magnetic coherence lengths of the order of tens of kpc, it is still not possible to unambiguously attribute these small-scale variations to the cluster alone.  One promising new way to measure the coherence of the cluster magnetic field unambiguously on such scales is through the use of so-called `cork screw' BTs such as ESO 137-G 007 in the Norma cluster (Figure \ref{fig:helix}; left panel). `Corkscrew' BTs result when the radio galaxy spirals as it falls into the cluster potential giving rise to long helical tails with narrow pitch angles. In such tails the front and back parts of the helix are probing two different path-lengths through the ICM which only differ by $\sim20$ kpc in cases such as shown here ($\Delta d$ in the right panel of Figure \ref{fig:helix}). This path difference between the near and far side of the jets should result in a two distinct Faraday rotation measure (RM) values, due to the extra RM component ($\Delta$RM) over the additional distance, $\Delta d$. This implies that statistical differences between the average rotation measures of patches associated with the front and back of the helical tails are likely to arise from within the cluster in which the galaxy is embedded.  Although rare such sources should be seen in large numbers in the future with instruments such as the Square Kilometre Array (SKA). We estimate that SKA1 should detect 5000 to 10,000 such sources in an all-sky survey. 

\section{Conclusion}

Recently we have seen the utility of searching for BTs in wide-area radio surveys such as FIRST, with of order 2000 sources detected (Wing \& Blanton 2011). Next generation radio telescopes are predicted to detect BTs in even greater numbers. In the case of ASKAP it is estimated that the all-sky continuum survey Evolutionary Map of the Universe (EMU, Norris et al. 2011) will detect just under half a million BTs out to a redshift of 2 and with sizes down to tens of kpc (Dehghan et al. 2014). For the SKA in phase 1, the likely number reaches 1 million (Johnston-Hollitt et al. 2015). Of these it has been estimated that 50,000 to 100,000 will be sufficiently polarised to undertake rotation measure studies within galaxy clusters (Johnston-Hollitt et al. 2015) and here we estimate that of those, 5,000 to 10,000 will be the so-called `corkscrew' BTs which can be used to probe the magnetic field coherence lengths in clusters along the line-of-sight on scales of tens of kpc. Even if these instruments do not achieve the currently planned sensitivities, it is clear that vast numbers of BTs will be detected, signalling the location of new clusters, groups and dense regions in the Universe.  The use of these objects as probes to examine their environment will thus accelerate, providing critical information on cluster densities, winds and velocity flows over a large sample of systems of varying mass. Additionally, we may perform the first unambiguous constraint on turbulence in the intracluster medium on scales which have thus far not been possible.


\newpage

\begin{thebibliography}{}

\bibitem[Blanton \etal\ (2001)]{Blanton_etal01}
{Blanton, E. L., Gregg, M. D., Helfand, D. J., et al.} 2001,
\textit{ApJ}, 121, 2915

\bibitem[Blanton \etal\ (2003)]{Blanton_etal03}
{Blanton, E. L., Gregg, M. D., Helfand, D. J., et al.} 2003,
\textit{ApJ}, 125, 1635

\bibitem[Blanton \etal\ (2015)]{Blanton_etal15}
{Blanton, E. L., Paterno-Mahler, R., Wing, J. D., \etal} 2015, these proceedings

\bibitem[Dehghan \etal\ (2011)]{Dehghan_etal11}
{Dehghan, S., Johnston-Hollitt, M., Mao, M., et al.} 2011,
\textit{JA\&A}, 32, 491

\bibitem[Dehghan\ (2014)]{Dehghan14}
{Dehghan, S.} 2014,
\textit{PhD Thesis}, Victoria University of Wellington

\bibitem[Dehghan \etal\ (2014)]{Dehghan_etal14a}
{Dehghan, S. \& Johnston-Hollitt, M.} 2014,
\textit{ApJ}, 147, 52

\bibitem[Dehghan \etal\ (2014)]{Dehghan_etal14b}
{Dehghan, S., Johnston-Hollitt, M., Franzen, T. M. O., et al.} 2014,
\textit{AJ}, 148, 75

\bibitem[Dehghan \etal\ (2015)]{Dehghan_etal15}
{Dehghan, S.,  Johnston-Hollitt, M. et. al} 2015, in preparation

\bibitem[Douglass \etal\ (2011)]{Douglass_etal11}
{Douglass, E. M., Blanton, E. L., Clarke, T. E., et al.} 2011,
\textit{ApJ}, 743, 199

\bibitem[Freeland \etal\ (2008)]{Freeland_etal08}
{Freeland, E., Cardoso, R. F., \& Wilcots, E.} 2008,
\textit{ApJ}, 685, 858

\bibitem[Johnston-Hollitt \etal\ (2015)]{Johnston-Hollitt_etal15}
{Johnston-Hollitt, M., Dehghan, S., \& Pratley, L.} 2015,
\textit{Proceedings of ``Advancing Astrophysics with the Square Kilometre Array", PoS (AASKA14)}, paper 101

\bibitem[Mao \etal\ (2009)]{Mao_etal09}
{Mao, M.Y., Johnston-Hollitt, M.,Stevens, J.B., \& Wotherspoon, S. J.} 2009,
\textit{MNRAS}, 392, 1070

\bibitem[Mao \etal\ (2010)]{Mao_etal10}
{Mao, M.Y., Sharp, R., Saikia, D.J., et al.} 2010,
\textit{MNRAS}, 406, 2578

\bibitem[Norris \etal\ (2011)]{Norris_etal11}
{Norris, R. P., \etal} 2011,
\textit{PASA}, 28, 215

\bibitem[Pfrommer \& Jones\ (2014)]{Pfrommer_etal14}
{Pfrommer, C. \& Jones, T. W.} 2011,
\textit{ApJ}, 730, 22

\bibitem[Pratley \etal\ (2013)]{Pratley_etal13}
{Pratley, L., Johnston-Hollitt, M., Dehghan, S., \& Sun M.} 2013,
\textit{MNRAS}, 432, 243 

\bibitem[Pratley \etal\ (2014)]{Pratley_etal15}
{Pratley, L., Johnston-Hollitt, M., Dehghan, S., \& Sun M.} 2015, these proceedings

\bibitem[Wing \& Blanton\ (2011)]{Blanton11}
{Wing, J. D. \& Blanton, E. L.} 2011,
\textit{AJ}, 141, 88

\end{thebibliography}
\end{document}